 \documentclass[aps,superscriptaddress,twocolumn,prb]{revtex4}
 \newcommand{\bc}{\begin{center}}
 \newcommand{\ec}{\end{center}}

 \newcommand{\be}{\begin{equation}}
 \newcommand{\ee}{\end{equation}}
 \renewcommand{\baselinestretch}{1.5}
 \usepackage{graphicx}
 \usepackage{epsfig}

 \def\e0{{e_0}}

  \def\be{\begin{equation}}
  \def\ee{\end{equation}}
  \def\bea{\begin{eqnarray}}
  \def\eea{\end{eqnarray}}

  \def\beq{\begin{equation}}
  \def\eeq{\end{equation}}
  \def\beqn{\begin{eqnarray}}
  \def\eeqn{\end{eqnarray}}
  
  \def\bc{\begin{center}}
  \def\ec{\end{center}}
  
  \def\etal{{\it et al}}

  \newcommand{\singlespace}{
          \renewcommand{\baselinestretch}{1}
          \large\normalsize}
  \newcommand{\doublespace}{
          \renewcommand{\baselinestretch}{1.75}
          \large\normalsize}
 \doublespace
 \singlespace
\begin{document}
  \title{Scattering of a Baseball by a Bat }

 \author{Rod Cross}\email{cross@physics.usyd.edu.au}
 \affiliation{\mbox{Physics Department, University of Sydney, Sydney NSW 2006, Australia}}

 \author{Alan M. Nathan}\email{a-nathan@uiuc.edu}
 \affiliation{\mbox{Department of Physics, University of Illinois, Urbana, IL 61801}}

  %\vspace{0.25in}
 % DRAFT
 \date{\today}

 \begin{abstract}
A ball can be hit faster if it is projected without spin but it can be hit farther if it is projected with backspin.
Measurements are
presented in this paper of the tradeoff between speed and spin for a baseball impacting a baseball bat.  The
results are inconsistent with a collision model in which the ball rolls off the bat and instead imply
tangential compliance in the ball, the bat, or both.  If the results are extrapolated to the higher speeds
that are typical of the game of baseball, they suggest
that a curveball can be hit with greater backspin than a fastball, but by an amount that
is less than would be the case in the absence of tangential compliance.
 \end{abstract}

 \maketitle

 \section{Introduction}

 Particle scattering experiments have been conducted for many years to probe the
 structure of the atom, the atomic nucleus, and the nucleons. By comparison, very few
 scattering experiments have been conducted with macroscopic objects. In this paper we
 describe an experiment on the scattering of a baseball by a baseball bat, not to probe the
 structure of the bat but to determine how the speed and spin of the outgoing ball depends on
 the scattering angle. In principle the results could be used to determine an appropriate force
 law for the interaction, but we focus attention in this paper on directly observable parameters.
 The main purpose of the experiment was to determine the amount of backspin that can be imparted to a
 baseball by striking it at a point below the center of the ball. The results are of a preliminary nature
 in that they were obtained at lower ball speeds than those encountered in the field. As such, the experiment
 could easily be demonstrated in the classroom or repeated in an undergraduate laboratory as an introduction
 to scattering problems in general.

 A golf ball is normally lofted with backspin so that the aerodynamic lift force will carry the ball as far as possible.
 For the same reason, a baseball will also travel farther if it is struck with backspin. It also travels farther if
 it is launched at a higher speed. In general there is a tradeoff between the
spin and speed that can be imparted to a ball, which is affected
in baseball by the spin and speed of the pitched ball. Sawicki,
\etal.\cite{Sa03}, henceforth referred to as SHS, examined this
problem and concluded that a curveball can be batted farther than
a fastball despite the higher incoming and outgoing speed of the
fastball.  The explanation is that a curveball is incident with
topspin and hence the ball is already spinning in the correct
direction to exit with backspin. A fastball is incident with
backspin so the spin direction needs to be reversed in order to
exit with backspin. As a result, the magnitude of the backspin
imparted to a curveball is larger than that imparted to a fastball
for any given bat speed and impact point on the bat, even allowing
for the lower incident speed of a curveball.  According to
SHS,\cite{Sa03} the larger backspin on a hit curveball more than
compensates the smaller hit ball speed and travels farther than a
fastball, a conclusion that has been challenged in the
literature.\cite{Ad05}
\newline
\indent SHS\cite{Sa03} assumed that a batted ball of radius $r$
will roll off the bat with a spin $\omega$ given by $r\omega =
v_{x}$ where $v_{x}$ is the tangential speed of the ball as it
exits the bat. However, a number of recent
experiments\cite{Cr02,Cr02a,Cr03,Cr03a,Cr05} have shown that balls
do not roll when they bounce. Rather, a ball incident obliquely on
a surface will grip during the bounce and usually bounces with
$r\omega > v_{x}$ if the angle of incidence is within about
$45^{\circ}$ to the normal. The actual spin depends on the
tangential compliance or elasticity of the mating surfaces and is
not easy to calculate accurately. For that reason we present in
this paper measurements of speed, spin and rebound angle of a
baseball impacting with a baseball bat. The implications for
batted ball speed and spin are also described.
\section{Experimental procedures}

 A baseball was dropped vertically onto a stationary, hand--held
 baseball bat to determine the rebound speed and spin as functions
 of (a) the scattering angle and (b) the magnitude and direction of
 spin of the incident ball. The impact distance from the
 longitudinal axis of the bat was varied on a random basis in order
 to observe scattering at angles up to about $120^{\circ}$ away
 from the vertical. Measurements were made by filming each bounce
 with a video camera operating at 100 frames/s, although
 satisfactory results were also obtained at 25 frames/s. The bat
 chosen for the present experiment was a modified Louisville Slugger model
 R161 wood bat  of length 84 cm (33 in.) with a barrel diameter of
 6.67 cm (2 $\frac{5}{8}$ in.) and mass $M = 0.989$ kg (35 oz).
 The center of mass of the bat was located 26.5~cm from the barrel
 end of the bat.  The moments of inertia about axes
 through the center of mass and perpendicular and parallel, respectively, to the longitudinal axis of the bat
 were 0.0460 and 4.39 x 10$^{-4}$ kg-m$^2$.
 The ball was a Wilson A1010, having a mass
 0.145~kg and diameter 7.2~cm.

 The bat was held in a horizontal position by one hand and the ball
 was dropped from a height of about 0.8~m using the other hand. A
 plumb bob was used to establish a true vertical in the video image
 and to help align both the bat and the ball. The ball was dropped
 either with or without spin. In order to spin the ball, a strip of
 felt was wrapped around a circumference and the ball was allowed
 to fall vertically while holding the top end of the felt strip.  A
 line drawn around a circumference was used to determine the ball
 orientation in each frame in order to measure its spin.  The
 impact distance along the axis was determined by eye against marks
on the barrel to within about 5~mm. If the ball landed 140 to 160
 mm from the barrel end of the bat the bounce was accepted. Bounces
 outside this range were not analyzed.

 The velocity of the ball immediately prior to and after impact was determined to within 2\% by
 extrapolating data from at least three video frames before and after each impact.  The horizontal
 velocity was obtained from linear fits to the horizontal coordinates of the ball and the vertical
 velocity was obtained from quadratic fits assuming a vertical acceleration of 9.8~m/s$^2$.
 Additional measurements were made by bouncing the ball on a hard wood floor to determine the normal
 and tangential COR, the latter defined below in Eq.~\ref{eq:ex},
and a lower limit on the coefficient of sliding friction ($\mu_k$)
between the ball and the floor. The COR values were determined by
dropping the ball with and without spin from a height of about
1.5~m to impact the floor at a speed of $5.6 \pm 0.3~$m/s.  The
incident ball spin was either zero, $-72 \pm 2$~rad/s or $+68 \pm
3$~rad/s. The normal COR $e_y$ was $0.59 \pm 0.01$, and the
tangential COR $e_x$ was $0.17 \pm 0.03$, corresponding to a
rebound spin $\omega_{2} \approx 0.16\omega_{1}$, where
$\omega_{1}$ is the incident spin. If a spinning baseball is
dropped vertically onto a hard floor, then it would bounce with
$\omega_{2}  = 0.29\omega_{1}$ if $e_{x} = 0$ (as assumed by
SHS\cite{Sa03}). The lower limit on $\mu_k$ was determined by
throwing the ball obliquely onto the floor at angles of incidence
between $25^{\circ}$  and $44^{\circ}$ to the horizontal, at
speeds from 3.5 to 4.2~m/s and with negligible spin. The value of
$\mu_k$ was
found from the data at low angles of incidence to be larger than
$0.31 \pm 0.02$. At angles of incidence between $30^{\circ}$ and
$44^{\circ}$ the ball did not slide throughout the bounce but
gripped the floor during the bounce, with $e_{x} = 0.14 \pm 0.02$.

 \section{Theoretical bounce models}

 \begin{figure}[htb]
 \epsfig{file=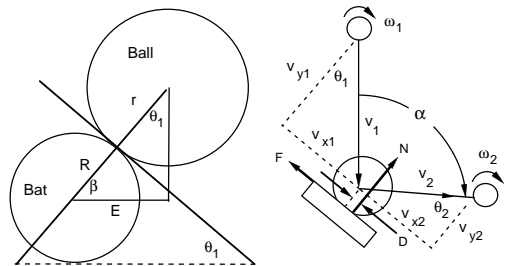,width=3in}
 \caption {Bounce geometry for a baseball of radius $r$, mass $m$ falling vertically onto a bat of
 radius $R$, mass $M$ with impact parameter $E$.}
 \label{fig:geometry}
 \end{figure}

 Consider the situation shown in Fig.~\ref{fig:geometry} where a ball of radius $r$ falls vertically
 onto a bat of radius $R$. In a low speed collision the bat and the ball will remain approximately
 circular in cross section. If the impact parameter is $E$, then the line joining the bat and ball
 centers is inclined at an angle $\beta$ to the horizontal where cos$\, \beta = E/(r + R)$. The ball
 is incident at an angle $\theta_{1} = 90 - \beta$ to the line joining the bat and ball centers and
 will rebound at an angle $\theta_{2}$. The ball is therefore scattered at an angle $\alpha = \theta_{1} + \theta_{2}$.
 During the collision, the ball experiences a tangential force $F$ and a normal
 force $N$.  For the low-speed collisions investigated here, the ball-bat force acts essentially at a point, so that
 the angular momentum of the ball about that point is conserved.  Indeed, low-speed collisions of tennis balls are consistent with
 angular momentum conservation.\cite{Cr02a}  However, high-speed collisions of tennis balls are known not to conserve angular
 momentum.\cite{Cr03}
 A phenomenological way to account for
 non-conservation of angular momentum is to assume that the normal force $N$ does not act through the center of
 mass of the ball but is displaced from it by the distance $D$,\cite{Cr02a} as shown in
 Fig.~\ref{fig:geometry} and discussed more fully below.

 The collision is essentially equivalent to one between a ball and a plane surface inclined at angle
 $\theta_{1}$ to the horizontal. Suppose that the ball is incident with angular velocity $\omega_{1}$
 and speed $v_{1}$. Let $v_{y1} = v_{1} \, {\rm cos} \, \theta_{1}$ denote the component of the incident
 ball speed normal to the surface and $v_{x1} = v_{1} \, {\rm sin} \, \theta_{1}$ denote the tangential
 component. The ball will bounce at speed $v_{y2}$ in a direction normal to the surface, with tangential
 speed $v_{x2}$ and angular velocity $\omega_{2}$. If the bat is initially at rest, it will recoil with
 velocity components $V_{y}$ and $V_{x}$ respectively perpendicular and parallel to the surface, where the
 velocity components refer to the impact point on the bat. The recoil velocity at the handle end or the
 center of mass of the bat are different since the bat will tend to rotate about an axis near the end of the handle.

 The bounce can be characterized in terms of three independent parameters:
 the normal coefficient of restitution (COR)
 $e_{y} = (v_{y2} - V_{y})/v_{y1}$; the tangential COR, $e_{x}$, defined by
 \be
e_{x} = - \frac{(v_{x2} - r\omega_{2} - [V_{x} -
R\Omega])}{(v_{x1}-r\omega_{1})}
 \label{eq:ex}
 \ee
 where $\Omega$ is the angular velocity of the bat about the longitudinal axis immediately after the
 collision; and the parameter $D$.
 The two coefficients of restitution are defined respectively in terms of the normal and tangential speeds of the
 impact point on the ball, relative to the bat, immediately after and immediately before the bounce.
 The bounce can also be characterized in terms of apparent coefficients of restitution, ignoring recoil
 and rotation of the bat. That is, one can define the apparent normal COR $e_{A} = v_{y2}/v_{y1}$ and the
 apparent tangential COR, $e_{T}$, given by
 \be
 e_{T} = - \frac{(v_{x2}-r\omega_{2})}{(v_{x1} - r\omega_{1})}
 \label{eq:eT}
 \ee
 There are three advantages of defining apparent COR values in this manner. The first is that apparent COR
 values are easier to measure since there is no need to measure the bat speed and angular velocity before or
 after the collision (provided the bat speed is zero before the collision). The second advantage is that the
 batted ball speed can be calculated from the measured apparent COR
 values for any given initial bat speed simply by a change of reference frame. We show how this is done in
 Appendix B. The third advantage is that the algebraic solutions of the collision equations are considerably
 simplified and therefore more easily interpreted.  Apparent and actual values of the COR are related by the expressions
 \be
 e_{A} = \frac{(e_{y} - r_y)}{(1 + r_y)}
 \label{eq:eAey}
 \ee
 and
 \be
 e_T=\frac{(e_x-r_x)}{(1+r_x)} + \frac{5}{2}
\left [\frac{D}{r}\right]\left (\frac{r_x}{1+r_x}\right
)\frac{v_{y1}(1+e_A)}{(v_{x1}-r\omega_1)}\, ,
 \label{eq:eTex}
 \ee
 where the recoil factors, $r_y$ and $r_x$, are the ratios of effective ball to bat masses for
 normal and tangential collisions, respectively.  An expression for $r_y$ was derived by Cross:\cite{Cross2000}
 \be
 r_y \,= \, m\left ( \frac{1}{M}\, +  \frac{b^2}{I_0}\right ) \,
 \label{eq:ry}
 \ee
 and an expression for $r_x$ is derived in Appendix A:
 \be
 r_x  \,=\, \frac{2}{7}m \left ( \frac{1}{M}\, +  \frac{b^2}{I_0}\,   \, + \, \frac{R^2}{I_z} \right
 )\, .
 \label{eq:rx}
 \ee
 In these expressions, $m$ is the ball mass, $I_0$ and $I_z$ are the moments of inertia (MOI) about an axis
 through the center of mass and  perpendicular and parallel, respectively, to the longitudinal axis of the bat,
 and $b$ is the distance parallel to the longitudinal axis between the impact point and the center of mass.
 For the bat used in the experiments at an impact distance 15~cm from the barrel end of the bat,
 $r_y$=0.188 and $r_x$=0.159, assuming the bat is free at both ends. The exit
 parameters of the ball are independent of
 whether the handle end is free or hand--held, as described previously by the authors.\cite{Cr99,Na00}
 Eq.~\ref{eq:eTex} will not be used in this paper except for some comments in Sec.~\ref{sec:insights}
 and for comparison with SHS\cite{Sa03} who assumed in their calculations that $e_{x} = 0$ and $D=0$,
 implying $e_T=-0.14$ for our bat.
 As discussed more fully in the next section, we find better agreement with our data when $e_{T} = 0$.

 From the definition of the parameter $D$, the normal force exerts a torque resulting in a change in
angular momentum of the ball about the contact point given by

\begin{widetext}
 \be
 \left(I\omega_{2} + mrv_{x2}\right ) - \left(I\omega_{1} + mrv_{x1}\right ) =
 -D\int Ndt = -mD(1+e_A)v_{y1}
 \label{eq:angmom}
 \ee
 \end{widetext}
 where $I = \alpha m r^{2}$ is the moment of inertia of the ball about its center of mass. For a solid sphere,
 $\alpha = 2/5$, although Brody has recently shown that $\alpha\approx 0.378$ for a baseball.\cite{Brody05}
 Eqs.~\ref{eq:eT} and \ref{eq:angmom} can be solved to show
 that
 \begin{widetext}
 \be
 \frac{v_{x2}}{v_{x1}} =   \frac{(1 - \alpha e_{T})}{(1 + \alpha)} + \frac{\alpha(1 + e_{T})}{(1 + \alpha)}
 \left (\frac{r\omega_{1}}{v_{x1}} \right)
 -\frac{D(1+e_A)}{r(1+\alpha)}\left
 (\frac{v_{y1}}{v_{x1}}\right )
 \label{eq:vx2}
 \ee
 and
 \be
 \frac {\omega_{2}}{ \omega_{1}} =  \frac{(\alpha  - e_{T})}{(1 + \alpha)} + \frac{(1 + e_{T})}{(1 +
 \alpha)} \left ( \frac {v_{x1}}{r\omega_{1}} \right )
 -\frac{D(1+e_A)}{r(1+\alpha)}\left
 (\frac{v_{y1}}{r\omega_1}\right ) \, .
 \label{eq:w2}
 \ee
 \end{widetext}
Eqs.~\ref{eq:vx2} and \ref{eq:w2}, together with the definition of
$e_A$ give a complete description of the scattering process in the
sense that, for given initial conditions, there are three
observables,  $v_{y2}$, $v_{x2}$, and $\omega_2$, and three unknown
parameters, $e_A$, $e_T$, and $D$, that can be inferred from a
measurement of the observables.
 We have written Eq.~\ref{eq:vx2} and \ref{eq:w2} for the general
 case of nonzero $D$.  However, as we will show in the next section, the present
 data are consistent with $D=0$, implying conservation of
 the ball's angular momentum about the point of contact.
 The normal bounce speed of the ball is determined by $e_{A}$, while for $D\approx 0$ the spin and tangential bounce speed
 are determined by $e_{T}$ and $r\omega_{1}/v_{x1}$. Depending on the magnitude and
 sign of the latter parameter,  $v_{x2}$ and $\omega_{2}$ can each be positive, zero or
 negative.  Eqs.~\ref{eq:vx2} and \ref{eq:w2} are generalizations of
 equations written down by Cross\cite{Cr02a} for the special case of the ball impacting a massive
 surface and reduce to those equations when $e_A=e_y$ and
 $e_T=e_x$.

 \section{Experimental results and discussion}
\subsection{Determination of $e_T$}
\label{sec:eT}

\begin{figure}[htb]
 \epsfig{file=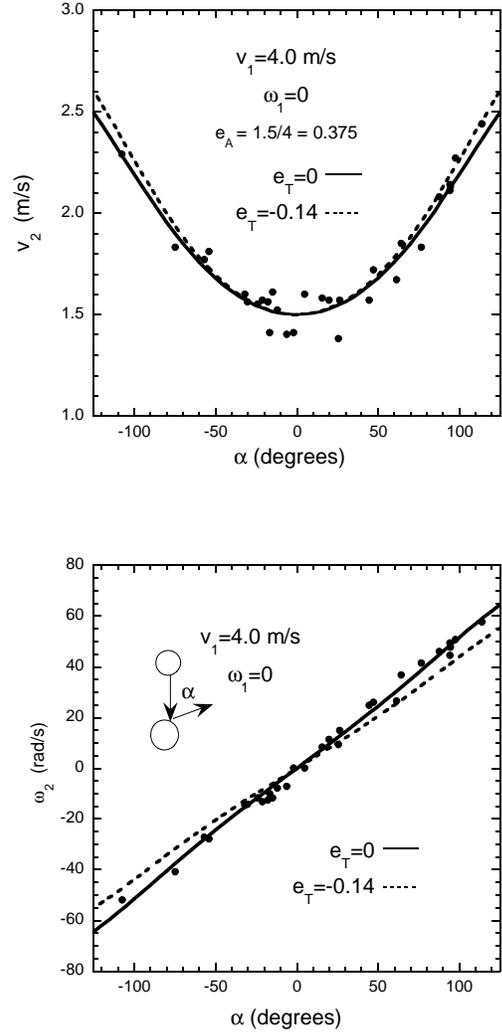,width=3in} \caption
 {Results with the ball incident with $\omega_{1} = 0$, along with theoretical
 curves calculated with $e_{T} = 0$ or -0.14} \label{fig:w1eq0}
 \end{figure}

\begin{figure}[tb]
 \epsfig{file=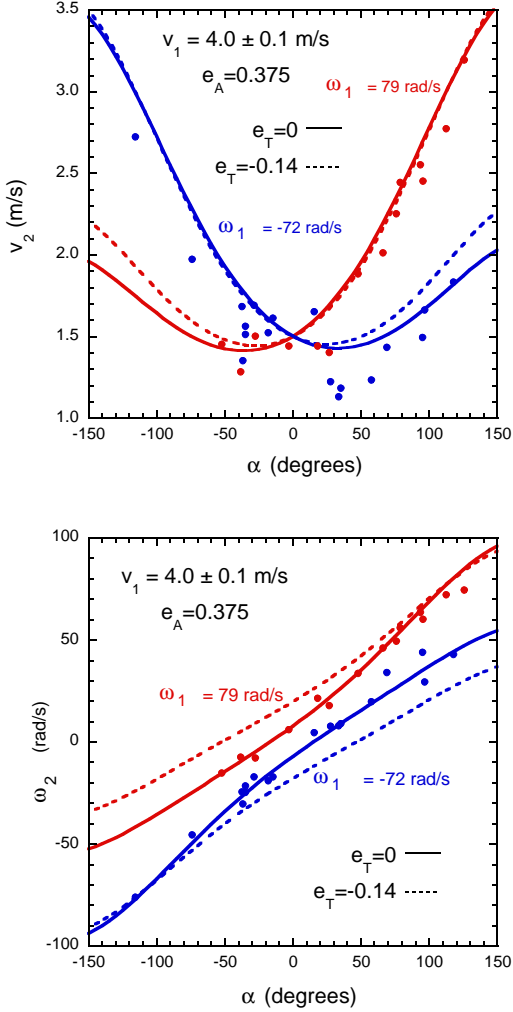,width=3in} \caption
 {Results with the ball incident with topspin or backspin, along with
 theoretical curves calculated with $e_{T} = 0$ or -0.14} \label{fig:w1ne0}
 \end{figure}

 We initially analyze the data using Eqs.~\ref{eq:vx2} and
\ref{eq:w2} assuming $D=0$ and postpone for the time being a
discussion of angular momentum conservation.  Results obtained when
the ball was incident on the bat without
 initial spin are shown in Fig.~\ref{fig:w1eq0}. The ball impacted
 the bat at speeds varying from 3.8 to 4.2~m/s but the
 results in Fig.~\ref{fig:w1eq0} were scaled to an incident
 speed of 4.0~m/s by assuming that the rebound speed and spin
 are both directly proportional to the incident ball speed, as
 expected theoretically.  An experimental value $e_{A} = 0.375 \pm
 0.005$  was determined from results at low (back) scattering angles,
 and this value was used together with Eqs.~\ref{eq:vx2} and \ref{eq:w2} to
 calculate rebound speed, spin, and scattering angle as functions of
 the impact parameter for various assumed values of $e_{T}$.  Best
 fits to the experimental data were found when $e_{T} = 0$ but
 reasonable fits could also be obtained with $e_{T} = 0 \pm 0.1$.
 Results obtained when the ball was incident with topspin or
 backspin are shown in Fig.~\ref{fig:w1ne0}. These results are not expected to
 scale with either the incident speed or incident spin and have not
 been normalized.  Consequently the data show slightly more scatter than
 those presented in Figs.~\ref{fig:w1eq0}. The ball impacted the bat at speeds varying from
 3.9 to 4.1~m/s and with topspin varying from 75 to
 83~rad/s or with backspin varying from -72 to
 -78~rad/s.  Simultaneous fits to all three data sets resulted in
 $e_{A} = 0.37 \pm 0.02$ and with $e_{T} = 0 \pm 0.02$.
 Using the recoil factors $r_y=0.188$ and $r_x=0.159$, our values for
 $e_A$ and $e_T$ imply $e_y=0.63 \pm 0.01$ and $e_x=0.16 \pm
 0.02$.  The result for $e_x$ is consistent with that measured by impacting the ball
 onto a hard floor ($0.17\pm0.03$) but the result for $e_y$ is slightly higher, presumably because of the
 lower impact speed and the softer impact on the bat.  On the other hand, Figs. \ref{fig:w1eq0} and \ref{fig:w1ne0} show that the
 measured $\omega_2$ values are inconsistent with $e_T=-0.14$,
 which is the result expected for our bat if $e_x=0$, as assumed by SHS.\cite{Sa03}

We next investigate the more general case in which angular momentum
is not conserved by fitting the data to Eqs.~\ref{eq:vx2} and
\ref{eq:w2} allowing both $D$ and $e_T$ as adjustable parameters.
Fitting to all three data sets simultaneously, we find $e_T=0\pm
0.02$ and $D=0.21\pm 0.29$ mm.  This justifies our earlier neglect
of $D$ and confirms that the data are consistent with angular
momentum conservation.  All calculations discussed below will assume
$D=0$.

\begin{figure}[htb]
 \epsfig{file=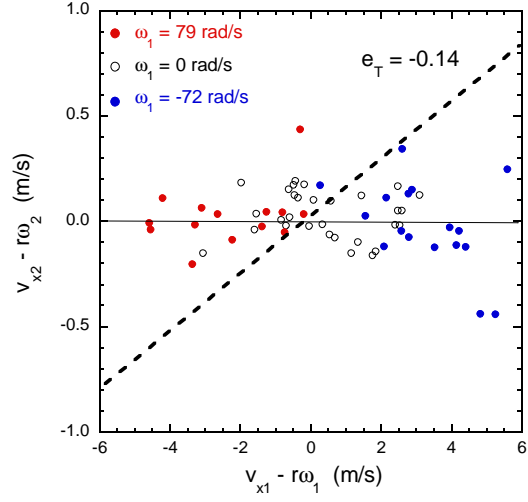,width=3in}
 \caption
 {Relationship between the final and initial tangential velocity of the ball.  For $e_T=0$, the final
 tangential velocity would be zero.  The dashed line is the expected result for $e_T=-0.14$.}
\label{fig:vt2vt1}
\end{figure}

 It is possible to determine the incident and outgoing angles with respect to the normal, $\theta_1$
 and $\theta_2$, from the measured quantities $v_1$, $v_2$, $\omega_1$, $\omega_2$, and $\alpha$ by
 applying angular momentum conservation about the contact point, Eq.~\ref{eq:angmom}, with $D$=0.
Once $\theta_1$ and $\theta_2$ are known, it becomes possible to
calculate other quantities of interest, such as the initial and
final tangential velocities $v_x-r\omega$.  The relationship
between these quantities is shown in Fig.~\ref{fig:vt2vt1}, where
it is seen that the final velocities are clustered around zero, as
would be expected for $e_T=0$ (see Eq.~\ref{eq:eT}).  When plotted
in this manner, it is quite clear that the data are completely
inconsistent with $e_T=-0.14$.  Note that the principal
sensitivity to $e_T$ comes from large values of
$|v_{x1}-r\omega_1|$.  This occurs whenever $v_{x1}$ and
$\omega_1$ have the opposite sign, leading to a reversal of the
spin and scattering angles that are negative for $\omega_1>0$ and
positive for $\omega_1<0$.  Indeed, Fig.~\ref{fig:w1ne0} shows
that those are exactly the regions of largest sensitivity to
$e_T$.

\subsection{Implications for batted balls}
\label{sec:implication}

We next explore the implications of our results for the spin and
speed of a batted ball, fully mindful that the present experiment
was done at very low speeds compared to those appropriate for the
game of baseball.  The goal of this analysis is not to make any
definite predictions about the spin and speed of a batted ball but
to examine the consequences of a small positive value of $e_x$
relative to the value $e_x=0$ assumed by SHS.\cite{Sa03}  Whether
such a value of $e_x$ is realized in a realistic ball-bat
collision will have to await experiments at higher speed.

\begin{figure}[htb]
\epsfig{file=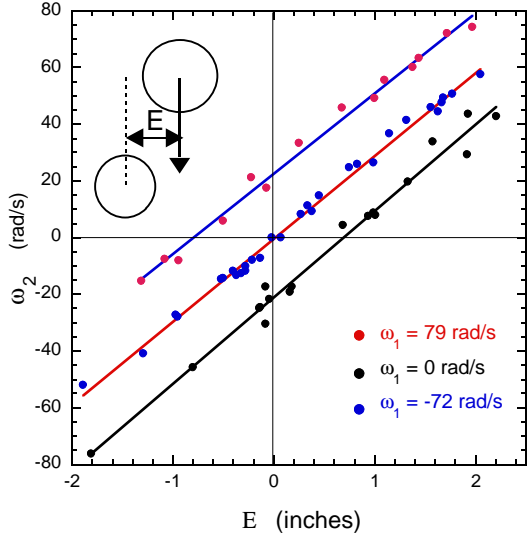,width=3in} \caption {Plot of the final
spin $\omega_2$ versus the impact parameter $E$, with $v_1\approx
4.0$ m/s.  These data clearly show that a ball with incident
topspin ($\omega_1 > 0$) has larger final spin than a ball with
incident backspin ($\omega_1 < 0$.)  The lines are linear fits to
the data.} \label{fig:wvsE}
\end{figure}

\begin{figure}[htb]
\epsfig{file=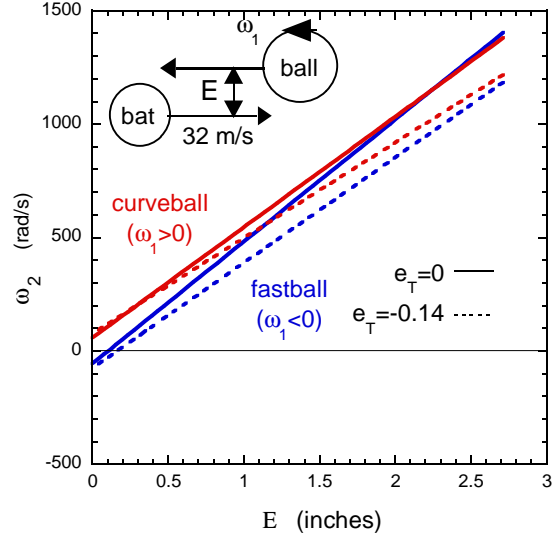,width=3in} \caption {Calculation of the
outgoing spin on a fastball ($\omega_1=-200$ rad/s, $v_1=42$ m/s)
and a curveball ($\omega_1=+200$ rad/s, $v_1=35$ m/s) for two
different values of $e_T$.} \label{fig:fbcb}
\end{figure}

With that caveat,  we first consider directly our own data in
Fig.~\ref{fig:wvsE}, where we plot $\omega_2$ versus the impact
parameter $E$, which is calculated from the inferred value of
$\theta_1$. These data demonstrate that for a given $E>0$, a ball
with initial topspin ($\omega_1>0$) has a larger outgoing backspin
than a ball with initial backspin, in qualitative agreement with
the argument of SHS.\cite{Sa03} To investigate the argument more
quantitatively, we apply our formalism to the ball-bat scattering
problem of SHS\cite{Sa03} to compare the final spin on a fastball
to that on a curveball. In this problem, the bat and ball approach
each other prior to the collision, thereby necessitating applying
a change of reference frame to the formulas already developed. The
relevant formulas, Eqs.~\ref{eq:vylab}--\ref{eq:w2lab}, are
derived in Appendix B. We assume that the initial velocity of the
bat is parallel to that of the ball, but displaced by the impact
parameter $E$, as shown in Fig.~\ref{fig:fbcb}.  The initial bat
speed is 32 m/s (71.6 mph). The incident fastball has a speed of
42 m/s (94 mph) and spin of -200 rad/s (-1910 rpm), whereas the
incident curveball has a speed of 35 m/s (78 mph) and a spin of
+200 rad/s. Using values of the normal COR $e_y$ assumed by
SHS\cite{Sa03,cor} the calculated final spin as a function of $E$
is presented in Fig.~\ref{fig:fbcb} both for $e_T=0$, as
determined from the present measurements, and for $e_T=-0.14$, as
assumed by SHS.\cite{Sa03}

Several important features emerge from this plot.  First, the
final spin $\omega_2$ is {\it less sensitive} to the initial spin
$\omega_1$ for $e_T=0$ than for $e_T=-0.14$.  This result is
consistent with Eq.~\ref{eq:w2}, where the first term on the
righthand side is larger for $e_T<0$ than for $e_T=0$. Our result
means that the difference in backspin between a hit fastball and
hit curveball is not as large as suggested by SHS.\cite{Sa03}
Second, the gap between the spin on the curveball and fastball
decreases as $E$ increases, a feature that can be understood from
the second term on the righthand side of Eq.~\ref{eq:w2}.  Since
the initial speed is larger for the fastball than the curveball,
the second term grows more rapidly for the fastball as $E$
increases, and eventually the two curves cross at $E\approx
2.4$~in. Moreover, the rate at which the two curves converge is
greater for $e_T=0$ than for $e_T=-0.14$. Had the initial speeds
been identical, the two curves would have been parallel. Third,
for $E\agt0.5$~in. and independent of the sign of $\omega_1$,
$\omega_2$ is {\it larger} when $e_T=0$ than when $e_T<0$, since
$\omega_2$ is mainly governed by the second term on the righthand
side of Eq.~\ref{eq:w2}.  The increase in $\omega_2$ is
accompanied by a decrease in $v_{x2}$, as required by angular
momentum conservation, and therefore by a slightly smaller
scattering angle. However, the outgoing speed is dominated by the
normal component, so the decrease in $v_{x2}$ hardly affects the
speed of the ball leaving the bat, at least for balls hit on a
home run trajectory.

These results will have implications for the issue of whether an optimally
hit curveball will travel farther than an optimally hit fastball.
To investigate this in detail requires a calculation
of the trajectory of a hit baseball, much as was done by SHS.\cite{Sa03}  Such
a calculation requires knowledge of the lift and drag forces on a spinning
baseball.  However, given the current controversy about these forces,\cite{Ad05}
further speculation on this issue is beyond the scope of the present work.

It is interesting to speculate on the relative effectiveness of
different bats regarding their ability to put backspin on a
baseball. As we have emphasized, the effectiveness is determined by
a single parameter, $e_T$, which in turn is related to the
horizontal COR $e_x$ and the recoil factor $r_x$.  For a given
$e_x$, a bat with a smaller $r_x$ will be more effective than one
with a larger $r_x$ (see Eq.~\ref{eq:eTex}).  Since $r_x$ is
dominated by the term involving $R^2/I_z$ (see Eq.~\ref{eq:rx}), one
might expect it to be very different for wood and aluminum bats.
Because of the hollow thin-walled construction of an aluminum bat,
it will have a larger MOI about the longitudinal axis ($I_z$) than a
wood bat of comparable mass and shape.  This advantage is partially
offset by the disadvantage of having a center of mass farther from
the impact point ($b$ is larger), which increases $r_x$. As a simple
numerical exercise, we have investigated two bats having the shape
of an R161, one a solid wood bat and the other a thin-walled
aluminum bat.  Both bats are 34~in. long and weigh 31.5 oz. The wood
bat has $I_0$=2578 and $I_z$=18.0 oz-in$^2$ and the center of mass
22.7 in. from the knob. The aluminum bat has $I_0$=2985 and
$I_z$=29.3 oz-in$^2$ and the center of mass 20.2 in. from the knob.
With an impact 6 in. from the barrel end, where the bat diameter is
2.625 in, and assuming $e_x$=0.16, then $e_T$ is -0.03 and 0,
respectively, for the wood and aluminum bat.  We conclude that,
generally speaking, an aluminum bat is marginally more effective in
putting backspin on the baseball than a wood bat of comparable mass
and shape.

\subsection{Insights into the scattering process}
\label{sec:insights}
 Besides the obvious practical implications of our result, it is interesting to ask what it teaches us
  about the scattering process itself.  As mentioned earlier, our measured value $e_T=0$ necessarily implies
   that $e_x\approx 0.16$.  A value $e_x<0$ would be obtained if the ball slides on the surface throughout the
    collision, whereas a value $e_x=0$ would be obtained if the ball is rolling when it leaves the bat.  However, a
     positive value of $e_x$ necessarily implies tangential compliance in the ball, the bat, or both. A rigid baseball
      impacting on a rigid bat without any tangential compliance in the contact region will slide on the bat until the
       contact point comes to rest, in which case it will enter a rolling mode and it will continue to roll with zero
        tangential velocity as it bounces off the bat.\cite{Br84}  However, a real baseball and a real bat can store energy elastically
         as a result of deformation in directions both perpendicular and parallel to the impact surface.  In that case, if
          tangential velocity is lost temporarily during the collision, then it can be regained from the elastic energy
           stored in the ball and the bat as a result of tangential deformation in the contact region. The ball will then
            bounce in an ``overspinning" mode with $r\omega_{2} > v_{x2}$ or with $e_{x} > 0$. The details of this process
             were first established by Maw, {\em et al}.\cite{Ma76,Ma81} The effect is most easily observed in the bounce
              of  a superball\cite{Ga69} which has tangential COR typically greater than
              0.5.\cite{Cr02,Cr02a} A simple lumped-parameter
              model for the bounce of a ball with tangential
            compliance has been developed by
            Stronge.\cite{Stronge00}

\begin{figure}[htb]
\epsfig{file=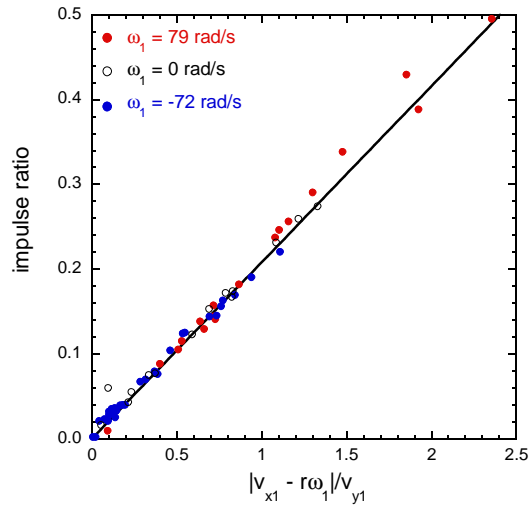,width=3in} \caption {The ratio of tangential to normal
impulse, Eq.\ref{eq:mumin}, plotted as a function of the
initial ratio of tangential to normal speed.  The line is the expected
impulse ratio for $e_A=0.375$ and $e_T=0$.} \label{fig:mumin}
\end{figure}

As mentioned above, the signature for continuous sliding throughout
the collision is $e_x<0$.  Referring to Fig.~\ref{fig:vt2vt1}, data
with $e_x<0$ would lie above or below the dashed line for values of
$v_{x1}-r\omega_1$ greater than or less than zero, respectively. Not
a single collision satisfies that condition in the present data set,
suggesting that $\mu_k$ is large enough to bring the sliding
to a halt.  Therefore the scattering data themselves can be used to set a {\it lower
limit} on $\mu_k$, which must be at least as large as the
ratio of tangential to normal impulse to the center of mass of the
ball:
\beq
\mu_k \, \geq \,\frac{\int Fdt}{\int Ndt} \, = \, \frac{v_{x2}-v_{x1}}{(1+e_A)v_{y1}} \, .
\label{eq:mumin}
\eeq
In Fig.~\ref{fig:mumin} values of the RHS of Eq.~\ref{eq:mumin}
are plotted as a function of the initial ratio of
tangential to normal speed.  Using the results derived in Appendix A, it is straightforward to show that
these quantities are linearly proportional with a slope equal to $(2/7)(1+e_T)/(1+e_A)$, provided the
initial velocity ratio is below the critical value needed to halt the sliding.  Stronge shows\cite{Stronge00} and
we confirm with our own formalism that the critical value is $(7/2)\mu_k(1+e_A)(1+r_x)$, which in our experiment
assumes the numerical value 5.6$\mu_k$.  Above the critical value,
the impulse ratio should be constant and equal to $\mu_k$.
Given that the data still follow a linear relationship up to an
initial velocity ratio of 2.4, corresponding to an impulse ratio of 0.50, we conclude
that $\mu_k\ge 0.50$.  Indeed, if the actual $\mu_k$ were as small as 0.50, the critical value of
the initial velocity ratio would be 2.8, which exceeds the maximum value in the
experiment.
The lower limit of 0.50 is larger
than the lower limit of 0.31 that we measured from oblique
collisions of a nonspinning ball with the floor.  For that
experiment, the angle with the horizontal needed to
achieve continuous slipping is less than 20$^\circ$, which is smaller
than our minimum angle of 25$^\circ$.
Although no attempt was made
to measure the ball-bat $\mu_k$ directly, our lower limit is consistent
with $\mu_k=0.50\pm 0.04$ measured by SHS.\cite{Sa03}

Finally, we remark on our finding that the scattering data are
consistent with $D\approx 0$, implying that the angular momentum
of the ball is conserved about the initial contact point.  At low
enough initial speed, the deformation of the ball will be
negligible, so that the ball and bat interact at a point and the
angular momentum of the ball is necessarily conserved about that
point. Evidentally, this condition is satisfied at 4 m/s initial
speed. It is interesting to speculate whether this condition will
continue to be satisfied at the much higher speeds relevant to the
game of baseball, where the ball experiences considerable
deformation and a significant contact area during the collision.
Simple physics consideratons\cite{Cr02a} suggest that it will not.
A ball with topspin incident at an oblique angle will have a
larger normal velocity at the leading edge than the trailing edge,
resulting in a shift of the line of action of the normal force
ahead of the center of mass of the ball ($D>0$).  A similar shift
occurs when brakes are applied to a moving automobile, resulting
in a larger normal force on the front wheels than on the back.
Such a shift has been observed in high-speed collisions of tennis
balls.\cite{Cr02a,Br02}  Whether a comparable shift occurs in
high-speed baseball collisions will have to be answered with
appropriate experimental data.

 \section{Summary and conclusions}

We have performed a series of experiments in which a baseball is
scattered from a bat at an initial speed of about 4 m/s.  For the
particular bat that was used in the experiment, we find the
horizontal apparent coefficient of restitution $e_T$ is consistent
with 0 and inconsistent with the value $-0.14$ expected if the
ball is rolling at the end of its impact.  These results necessarily
imply tangential compliance in the ball, the bat, or both.  We further find
that the data are consistent with conservation of angular momentum of the ball
about the contact point and with a coefficient of sliding friction between the ball and bat
larger than 0.50.  Our results suggest
that a curveball can be hit with greater backspin than a fastball, but by an amount that
is less than would be the case in the absence of tangential compliance.  We
note that since our investigations were done at low speed, one
must proceed with caution before applying them to the higher
speeds that are typical of the game of baseball.

 \appendix

 \section{Relationship between $e_T$ and $e_x$}

In this section, we derive for tangential collisions the
relationship, Eqs.~\ref{eq:eTex} and \ref{eq:rx}, between the
apparent COR $e_T$ and the COR $e_x$.  The derivation follows
closely that presented by Cross\cite{Cross2000} for the
relationship between $e_A$ and $e_y$. We first solve the simple
problem involving the collision of two point objects in one
dimension.  Object A of mass $m$ and velocity $v_1$ is incident on
stationary object B of mass $M$. Object A rebounds backwards with
velocity $v_2$ and object B recoils with velocity $V$.  Our sign
convention is that $v_1$ is always positive and $v_2$ is positive
if it is in the opposite direction to $v_1$.  The collision is
completely specified by the conservation of momentum \be \int Fdt
\,=\, m (v_2+v_1) \, = \, M V \, , \label{eq:pconserv} \ee and the
coefficient of restitution
$$
e \, \equiv \, \frac{v_2+V}{v_1} \, , \nonumber
$$
where $F$ is the magnitude of the force the two objects exert on each
other.
We define the apparent coefficient of restitution
\be
e_A \, \equiv \, \frac{v_2}{v_1} \, , \nonumber
\label{eq:eA1}
\ee
and we seek a relationship between $e_A$ and $e$.  Using Eq.~\ref{eq:eA1}, we write $e$ as
$$
e \, = \, e_A \, + \, \frac{V}{v_1} \, , \nonumber
$$
then use Eq.~\ref{eq:pconserv} to find
$$
\frac{V}{v_1} \, = \, (1+v_2/v_1)\frac{m}{M} \, , \nonumber
$$
from which we easily derive our desired expression
$$
e_A \, = \, \frac{e-m/M}{1+m/M} \, . \nonumber
$$

\begin{figure*}[htb]
\epsfig{file=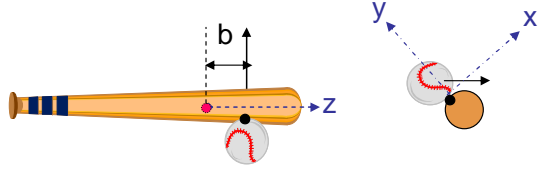,width=2.8in} \caption {Geometry for relating
$e_T$ and $e_x$.  The origin of the coordinate system is at the
center of mass of the bat, indicated by the red dot.  The $z$ axis
points along the long axis towards the barrel.  The $x$ and $y$
axes point, respectively, along the tangential and normal
directions. The solid arrow indicates the initial velocity of the
ball.  The black dot labels the point of contact P between ball
and bat.  In the right-hand figure, the $z$ axis points out of the
plane.} \label{fig:geom}
\end{figure*}

We next generalize this procedure for the collision of extended
objects, as shown in Figs.~\ref{fig:geometry} and \ref{fig:geom}.
Specifically we consider the collision between a ball and a bat.
 A ball of mass $m$ and radius $r$ is incident obliquely in the
 $xy$-plane on a stationary bat of mass $M$ and radius $R$ at the impact
 point.  Our
 coordinate system has its origin at the center of mass of the
 bat, with the $z$ axis along the longitudinal axis and the $x$
 and $y$ axes in the tangential and normal directions,
 respectively.  The impact point P has the coordinates $(0,R,b)$.
 The ball is incident with angular velocity $\omega_{1}$
 and linear velocity components $v_{x1}$ and $v_{y1}$; it rebounds with angular velocity
 $\omega_2$ and linear velocity $v_{x2}$ and $v_{y2}$, where the angular velocities are about the
 $z$ axis.    The bat recoils with CM velocity components $V_x$ and $V_y$
and with angular velocity about the CM with components $\Omega_x$, $\Omega_y$, and
$\Omega_z$.  Let $\vec{v}_{p1}$, $\vec{v}_{p2}$, and $\vec{V}_p$
denote the pre- and post-impact velocities of the ball and the post-impact velocity of the bat
at the point P, respectively.  Since we are concerned with tangential collisions, we only consider
the $x$ components of these velocities, which are given by
\bea
v_{p1x} \,&=&\, v_{x1}-r\omega_1 \nonumber \\
v_{p2x} \,&=&\, v_{x2}-r\omega_2 \nonumber \\
V_{px} \,&=&\, V_x+b\Omega_y+R\Omega_z
\label{eq:vp}
\eea
From the
definitions in Eq.~\ref{eq:ex} and \ref{eq:eT},
\bea
e_x \, &=& \,
- \frac{v_{p2x}-V_{px}}{v_{p1x}} \\ \nonumber
e_T \, &=& \, -
\frac{v_{p2x}}{v_{p1x}} \, . \nonumber
\eea
Applying the
impulse-momentum expressions to the bat, we find
\bea
\int Fdt \, &=& \, MV_x  \nonumber \\
b\int Fdt \, &=& \, I_0 \Omega_y \nonumber \\
R\int Fdt \, &=& \, I_z \Omega_z \, ,
\eea
where it has been
assumed that the bat is symmetric about the $z$ axis so that
$I_x=I_y\equiv I_0$. Combining these equations with
Eq.~\ref{eq:vp}, we find
\beq
\int Fdt \, = \, M_{ex}V_{px} \, ,
\eeq
where $M_{ex}$, the bat effective mass in the {\it x} direction, is
given by
\beq
\frac{1}{M_{ex}} \, = \, \frac{1}{M}\, +
\frac{b^2}{I_0}\,   \, + \, \frac{R^2}{I_z} \, .
\label{eq:Mex}
\eeq
Applying the impulse-momentum expressions to the ball we find
\bea
\int Fdt \, &=& \, -m(v_{x2}-v_{x1}) \nonumber \\
r\int Fdt -D\int Ndt \, &=& \, \alpha mr^2(\omega_2-\omega_1)
\eea
with $\alpha=2/5$ for a uniform sphere. Noting that
$\int Ndt=(1+e_A)mv_{y1}$ and combining the preceding equations with
Eq.~\ref{eq:vp}, we find
\beq
\int Fdt \, = \,
-m_{ex}(v_{p2x}-v_{p1x})+\frac{m_{ex}Dv_{y1}(1+e_A)}{r\alpha} \, ,
\eeq
where $m_{ex}$, the ball effective mass in the x direction, is
given by
\beq m_{ex} \, = \, \frac{\alpha}{1+\alpha}m \, .
\label{eq:mex}
\eeq Combining the equations for the bat and ball,
we arrive at
\beq
-m_{ex}(v_{p2x}-v_{p1x})+\frac{m_{ex}Dv_{y1}(1+e_A)}{r\alpha} \, = \,
M_{ex}V_{px} \, ,
\label{eq:pconserv1}
\eeq
which is analogous to
the momentum conservation equation for point bodies,
Eq.~\ref{eq:pconserv}, provided the velocities refer those at the
contact point P and the masses are effective masses.  Following
the derivation for point masses, we combined
Eq.~\ref{eq:pconserv1} with the definitions of $e_x$ and $e_T$ to
arrive at the result
\beq e_T \, = \,
\frac{e_x-m_{ex}/M_{ex}}{1+m_{ex}/M_{ex}}+ \left
[\frac{D}{r\alpha}\right]\left (\frac{r_x}{1+r_x}\right
)\frac{v_{y1}(1+e_A)}{(v_{x1}-r\omega_1)} \, .
\eeq
Defining
$r_x=m_{ex}/M_{ex}$ and assuming $\alpha=2/5$, then this equation,
along with Eq.~\ref{eq:Mex} and \ref{eq:mex}, is identical to
Eqs.~\ref{eq:eTex} and \ref{eq:rx}.  Our results are equivalent to
those used by SHS.\cite{Sa03}  We note that
Stronge\cite{Stronge00} has derived an expression that is
equivalent to Eq.~\ref{eq:rx} for the special case of a bat with
zero length, implying $b=0$, and $D=0$.  The present result is a
generalization of Stronge's expression.

\section{Collision formulas in the laboratory reference frame}

The formulas we have derived, Eqs.~\ref{eq:vx2}--\ref{eq:w2}, for
$v_{x2}$ and $\omega_2$ are valid in the reference frame in which
the bat is initially at rest at the impact point.  The usual (or
``laboratory'') frame that is relevant for baseball is the one where
both the bat and ball initially approach each other.  In this
section, we derive formulas for $v_{x2}$ and $\omega_2$ in the
laboratory frame.  Our coordinate system is the same as that shown
in Fig.~\ref{fig:geom}, where the $y$ axis is normal and the $x$
axis is parallel to the ball-bat contact surface.  In that system,
the initial velocity components of the bat and ball at the impact
point are denoted by $(V_x,V_y)$ and $(v_{x1}-r\omega_1,v_{y1})$,
respectively, where the usual situation has $V_y>0$ and $v_{y1}<0$.
In the bat rest frame, the components of the ball initial velocity
at the impact point are therefore
$(v_{x1}-r\omega_1-V_x,v_{y1}-V_y)$.  Applying the definitions of
$e_T$ and $e_A$, the components of the ball velocity after the
collision are given by
$$
v_{x2}-r\omega_2-V_x \, = \, - e_T\left (v_{x1}-r\omega_1-V_x \right )
$$
and
$$
v_{y2}-V_y \,= \, e_A\left (v_{y1}-V_y\right ) \, ,
$$
which can be rearranged to arrive at
\beq
v_{x2}-r\omega_2 \, = \, -e_T\left(v_{x1}-r\omega_1\right ) \, + \, (1+e_T)V_x
\label{eq:vtlab}
\eeq
and
\beq
v_{y2} \, = \, e_Av_{y1}\,+\,(1+e_A)V_y \, .
\label{eq:vylab}
\eeq
Finally, we combine Eq.~\ref{eq:vtlab} with the expression for angular momentum conservation, Eq.~\ref{eq:angmom}, to find the following
expressions for $v_{x2}$ and $w_2$:
%\begin{widetext}
\beq
 v_{x2} =   v_{x1}\frac{(1 - \alpha e_{T})}{(1 + \alpha)} +  (r\omega_1+V_x)\frac{\alpha(1 + e_{T})}{(1 +
 \alpha)}
 \label{eq:vx2lab}
 \eeq
 and
 \beq
 r\omega_2=  r\omega_1\frac{(\alpha  - e_{T})}{(1 + \alpha)} + (v_{x1}-V_x)\frac{(1 + e_{T})}{(1 +
 \alpha)}
              \,    .
 \label{eq:w2lab}
 \eeq
 %\end{widetext}
 Eqs.\ref{eq:vylab}--\ref{eq:w2lab} are the results we seek.
 Eq.~\ref{eq:vylab} has appeared in the literature many times.\cite{Br02,Nathan03}  To
 our knowledge, this is the first time anyone has written an explicit formula for $v_{x2}$ and $\omega_2$
 in the laboratory frame.  Although explicit formulas were not written down, the earlier works of
 SHS\cite{Sa03} and Watts and Baroni\cite{Watts} are equivalent to ours for the special case $e_x=0$, the latter being
 equivalent to $e_T=-0.137$ for our bat.
 Our formulas represent a generalization of their work to the case of
 arbitrary $e_x$.

%  \section*{References}
% \begin{references}


\begin{thebibliography}{}

 \bibitem{Sa03}  G.S. Sawicki, M. Hubbard and W.J. Stronge, ``How to hit home runs: Optimum baseball swing
 parameters for maximum range trajectories," Am. J. Phys. {\bf 71}, 1152--1162 (2003).

 \bibitem{Ad05} R. K. Adair, ``Comment on `How to hit home runs','' Am. J. Phys. {\bf 73}, 184--185 (2005);
 G. S. Sawicki, M. Hubbard, and W. J. Stronge, ``Reply to `Comment','' Am. J. Phys. {\bf 73}, 185--189 (2005).

 \bibitem{Cr02} R. Cross, ``Measurements of the horizontal coefficient of restitution for a superball and
 a tennis ball,'' Am. J. Phys. {\bf  70},482--489, (2002).

 \bibitem{Cr02a} R. Cross, ``Grip-slip behavior of a bouncing ball,'' Am. J. Phys. {\bf 70}, 1093-1102, (2002).

 \bibitem{Cr03} R. Cross, ``Measurements of the horizontal and vertical speeds of tennis courts,''
 Sports Engineering {\bf 6}, 95--111 (2003).

 \bibitem{Cr03a} R. Cross, ``Oblique impact of a tennis ball on the strings of a tennis racquet,''
 Sports Engineering {\bf 6}, 235--254 (2003).

 \bibitem{Cr05} R. Cross, ``Bounce of a spinning ball near normal incidence,'' Am. J. Phys. {\bf 73}, 914--920 (2005).

 \bibitem{Cross2000} R. Cross, ``The coefficient of restitution for collisions of happy ball, unhappy balls,
 and tennis balls,'' Am. J. Phys. {\bf 68}, 1025--1031 (2000).

 \bibitem{Cr99} R. Cross, ``Impact of a ball with a bat or racket,'' Am. J. Phys. {\bf 67}, 692--702, (1999).

 \bibitem{Na00} A. M. Nathan, ``Dynamics of the baseball--bat collision,'' Am. J. Phys. {\bf 68}, 979--990 (2000).


 \bibitem{Brody05} H. Brody, ``The moment of inertia of a tennis ball,'' Phys. Teacher {\bf 43}, 503--505 (2005).


 \bibitem{cor} The value of the normal COR depended somewhat on the relative normal ball-bat velocity and was
 approximately 0.435 for the fastball and 0.450 for the curveball.



 \bibitem{Br84}H. Brody, ``That's how the ball bounces,'' Phys. Teacher {\bf 22}, 494--497 (1984).


 \bibitem{Ma76}N. Maw, J. R. Barber, and J. N. Fawcett, ``The oblique impact of elastic spheres,'' Wear {\bf 38}, 101--114 (1976).

 \bibitem{Ma81}N. Maw, J. R. Barber, and J. N. Fawcett, ``The role of elastic tangential compliance in
 oblique impact,'' J. Lubrication Technology {\bf 103}, 74--80 (1981).

 \bibitem{Ga69}R. Garwin, ``Kinematics of an ultraelastic rough ball,'' Am. J. Phys. {\bf 37}, 88--92 (1969).

 \bibitem{Stronge00}  W. Stronge, {\it Impact Mechanics} (Cambridge University Press, Cambridge UK, 2000).

 \bibitem{Nathan03} A. M. Nathan, ``Characterizing the performance of baseball bats,'' Am. J. Phys. {\bf 71}, 134--143 (2003).

\bibitem{Br02} H. Brody, R. Cross and C. Lindsey. {\it The Physics and Technology of Tennis} (Racquet
Tech Publishing, Solana Beach, California, 2002).

  \bibitem{Watts} R. G. Watts and S. Baroni, ``Baseball-bat collisions and the resulting trajectories of
 spinning balls,'' Am. J. Phys.  {\bf 57}, 40--45 (1989).

%\end{references}

\end{thebibliography}
 \end{document}